\documentclass[paper]{ieice}
\usepackage[pdftex]{graphicx,xcolor}
\usepackage[fleqn]{amsmath}
\usepackage{newtxtext}
\usepackage[varg]{newtxmath}

\usepackage{xcolor}
\usepackage{balance}
\usepackage{booktabs}
\usepackage[caption=false,font=normalsize,labelfont=sf,textfont=sf]{subfig}
\usepackage{tikz}
\usepackage{cite}
\usepackage[belowskip=-10pt,aboveskip=0pt]{caption}
\usepackage{graphicx}

\newcommand{\AmSLaTeX}{%
 $\mathcal A$\lower.4ex\hbox{$\!\mathcal M\!$}$\mathcal S$-\LaTeX}
\def\BibTeX{{\rmfamily B\kern-.05em
 \textsc{i\kern-.025em b}\kern-.08em
  T\kern-.1667em\lower.7ex\hbox{E}\kern-.125emX}}
\makeatletter
\def\tmpcite#1{\@ifundefined{b@#1}{\textbf{?}}{\csname b@#1\endcsname}}%
\makeatother

\field{A}
\vol{107}
\no{1}
\SpecialSection{\LaTeXe\ Class File for the IEICE Transactions}
\title[]
      {A Corrugated All-Metal Vivaldi Antenna for 5G Phased Array Applications}

\authorlist{%
	\authorentry{Mahyar Mehri Pashaki}{n}{labelA}
	\authorentry{Mohammad Hossein Koohi Ghamsari}{n}{labelA}
	\authorentry{Alireza Mallahzadeh}{n}{labelA}
	\authorentry{Gabriele Gradoni}{n}{labelA}
	\authorentry{Mohsen Khalily}{n}{labelA}
}
\affiliate[labelA]{Institute for Communication Systems (ICS), Home of the 5G and 6G Innovation Centres (5G/6GIC), University of Surrey,
	Guildford GU2 7XH, U.K.
}

\begin{document}
\maketitle

\begin{summary}
In this paper, a corrugated Vivaldi phased array antenna in the 28 GHz frequency band is proposed for 5G communication applications. The presented configuration features an all-metal antipodal antenna structure with a broad bandwidth ranging from 26 to 30 GHz and beam steering capabilities from -30 to +30 degrees. The proposed antenna consists of a 4 $\times$ 4 array configuration, where each element has dimensions of 6.46 $\times$ 6.46 $\times$ 14.25 mm, resulting in an overall antenna structure with dimensions of 25.84 $\times$ 25.84 $\times$ 14.25 mm. The corrugation method is applied to minimize surface currents, resulting in a reduction in interelement mutual couplings. Therefore, the return loss in the array structure for central elements is decreased, and the antenna gain and radiation efficiency are improved. Moreover, the improved radiation efficiency allows for higher power transmission and reception from an antenna, resulting in potentially higher data rates and better performance.
\end{summary}
\begin{keywords}
5G, phased array, Vivaldi antenna, broadband antenna.
\end{keywords}

\section{Introduction}\label{intro}

The ever-growing demand for higher data rates in mobile communication systems, coupled with the advent of 5G and the Internet of Things (IoT), necessitates the development of antennas with improved gain characteristics \cite{yang2018digital}, \cite{chettri2019comprehensive}. However, increased antenna gain often comes at the expense of reduced coverage range, posing a challenge for mobile terminals and communication infrastructure.

One approach to addressing the coverage-gain trade-off is utilizing Millimeter-wave (mmWave) phased array antennas equipped with adaptive beamforming capabilities. These antennas are a cornerstone technology for achieving the high data rates and network capacity demanded by 5G communication systems \cite{ojaroudiparchin20168}, \cite{wang2022hybrid}. These arrays leverage beamforming to steer radio waves toward specific users electronically, enhancing signal quality and network efficiency compared to previous generations \cite{kahkonen2020surface}. 

In phased array antennas, the phase and amplitude of each element in the array are dynamically controlled to steer the radiation pattern toward the desired direction \cite{rivas2010review}. This enables targeted signal transmission and reception, significantly improving system capacity and spectral efficiency. Additionally, 5G phased arrays offer cost and power advantages by enabling spatial multiplexing, where a single base station serves multiple users simultaneously on the same frequency in non-overlapping spatial directions \cite{yang2017compact}, \cite{ibrahim2022slotted}. Moreover, recent advancements in integrated circuits and digital signal processing (DSP) have facilitated the development of high-performance, low-cost, and compact phased array systems, making them increasingly attractive for deployment in 5G and beyond-5G (B5G) applications \cite{sadhu20177}, \cite{sun2023reconfigurable}.

In recent years, several antenna configurations have been proposed as candidates for mmWave phased array antenna for wideband 5G communication applications. Specifically, Vivaldi antennas attracted great attention owing to their broad frequency range and straightforward construction. However, achieving radiation enhancement, reducing the mutual coupling effects, and increasing the effective bandwidth is difficult in the Vivaldi Antenna design due to its compact size and close element distances, necessitating a multifaceted strategy \cite{liu2017ultra}.

Various types of Vivaldi antenna arrays have been investigated in recent years \cite{gou2022radar,pan2020dual, kahkonen2019dual, azari2023super}. For example, \cite{gou2022radar} proposes low-RCS wideband Vivaldi phased arrays using array-level cancellation. In this work, two nearly identical Vivaldi elements with a 180° reflection phase difference are designed and used in a 16x16 array. Simulations showed a 10 dB average RCS reduction (6-18 GHz) compared to a reference array. A 30 GHz dual-polarized modular Vivaldi antenna array with integrated electronics for beam steering is presented in \cite{kahkonen2022modular}. The design achieves uniform amplitude and phase distribution across the array aperture with an adjustable radiation pattern and achieves a total gain between 35.6 dBi and 39.5 dBi.

This paper employs a corrugated technique and an antipodal structure to design a broadband, wide-angle beam-scanning Vivaldi phased array antenna that operates at the central frequency of 28 GHz. The corrugated technique involves increasing the effective electrical length of the horn antenna by implementing a periodic corrugation pattern on the interior surface of the horn. The following section presents the design and analysis procedure of the proposed Vivaldi antenna, along with the simulation results. The conclusions are also presented in section III.

\section{Simple Unit Element Design}
\begin{figure}[t]
	\centering
	
	$\begin{array}{cc}
		
		\includegraphics[scale=0.45,trim=10cm 3cm 14cm 1cm,clip=true]{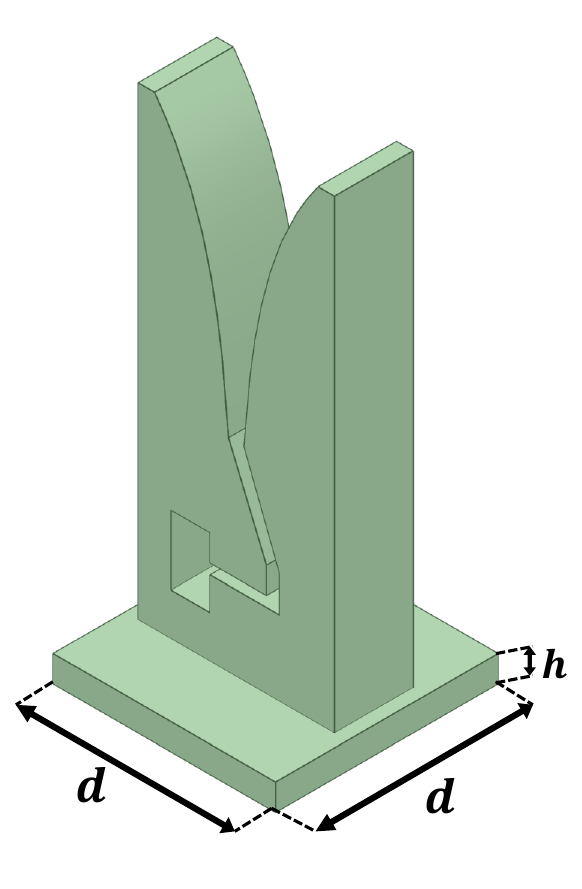}
		\includegraphics[scale=0.39,trim=11cm 1.5cm 5cm 0cm,clip=true]{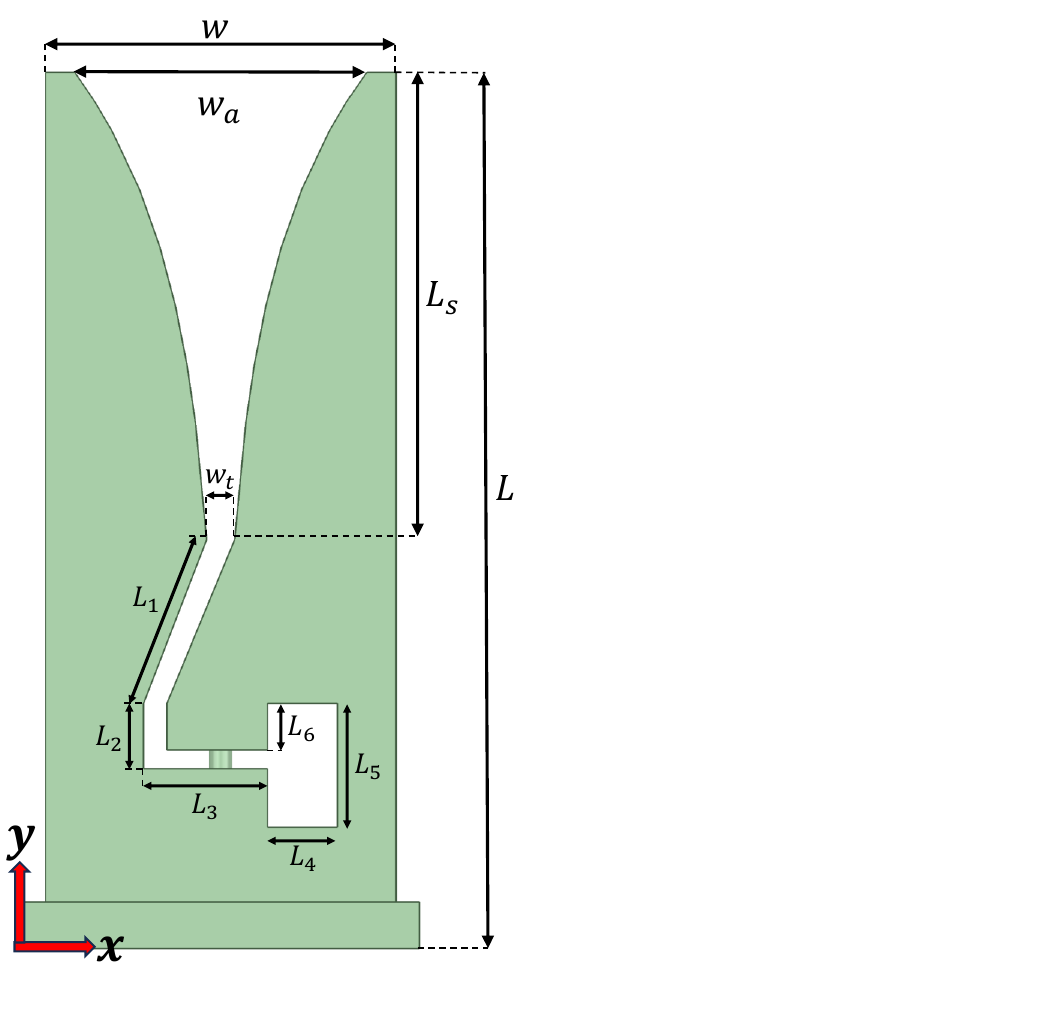}
		
	\end{array}$
	
	\captionsetup{
		font=footnotesize, 
		singlelinecheck=false, 
		margin={0pt,0pt} 
	}
	\caption{General geometry of a simple all-metal Vivaldi (SAMV) antenna unit element.}
	\label{fig1}
\end{figure}

The general structure of a simple all-metal Vivaldi (SAMV) antenna unit element is depicted in Fig.~\ref{fig1}. The dimensions of the lower rectangular plate are defined as $d=6.46\ mm$ and $h= 2\ mm$. Also, the thickness of the upper body is $t=2.28\ mm$ and the width of the aperture is $w=5.7\ mm$. The slot exponential curve, characterized by $w_t=0.45\ mm$, $w_a=4.75\ mm$, and $L_s=7.6\ mm$, determines the Vivaldi antenna radiation characteristics and impedance bandwidth. The tapering slot of an antenna facilitates the seamless transition from the feed impedance to the wave impedance. Additionally, reduced losses, enhanced radiation efficiency, and optimized bandwidth can be achieved with tapered slots. Implementing tapered openings can further reduce the size of the antenna, allowing it to occupy more confined spaces. The dimensions of the narrow slots defined after the aperture curve, $L_1$ to $L_6$, are $L_1=2.66\ mm$, $L_2= 1\ mm$, $L_3=2\ mm$, $L_4=1.15\ mm$, $L_5=2\ mm$, and $L_6= 0.8\ mm$ respectively.

The groove lines are engraved on the metal plate with the exponential slope to attain the broad frequency range anticipated from the Vivaldi antenna. In order for the aperture to radiate outward or receive the electromagnetic wave inward, the aperture must open progressively. A slot length ($L_s$) that is too short will produce an antenna with a limited bandwidth, whereas a slot length that is too long will produce a broadband antenna but with a decreased gain. To design an antenna with beam steering capability up to 70 degrees, we select $L_s$ value approximately equal to $\lambda/4$  and constrain the dimensions of each antenna element to $\lambda/2$. The slot curve profile of the antenna aperture can be expressed as (\ref{eq1}):

\begin{equation}
	y =  - (\frac{w_a}{2}){exp^{((\frac{1}{L_s})Log_{e}(\frac{w_t}{w_a})x)}} \label{eq1}\\
\end{equation}

\begin{figure}[!t]
	\centering
	
	\includegraphics[scale=0.6,trim=3cm 8cm 0cm 8cm,clip=true]{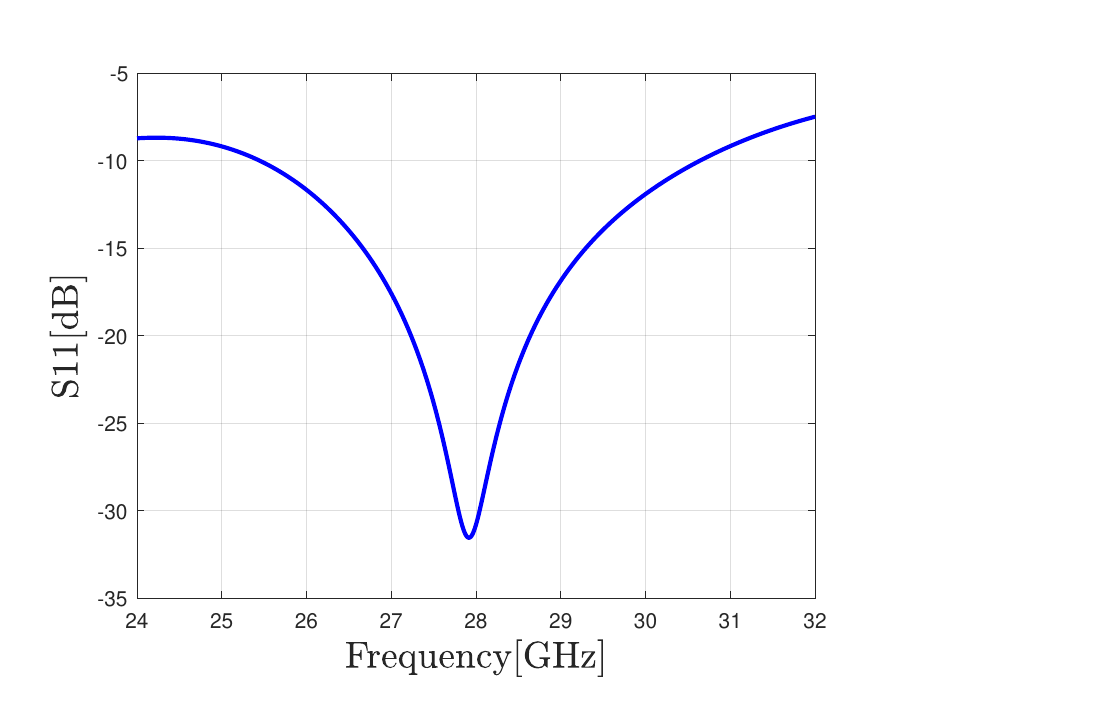}
	
	\captionsetup{
		font=footnotesize, 
		singlelinecheck=false, 
		margin={0pt,0pt} 
	}	
	\caption{Return loss (S11) of the SAMV antenna unit element.}
	\label{fig2}
\end{figure}

The impedance matching of the antenna 50-ohm port to the impedance of free space has been accomplished effectively through the use of appropriate tapering at the aperture. The return loss of the SAMV antenna unit element is shown in Fig.~\ref{fig2}.

\begin{figure}[!t]
	\centering
	
	\includegraphics[scale=0.05,trim=60cm 20cm 60cm 9.5cm,clip=true]{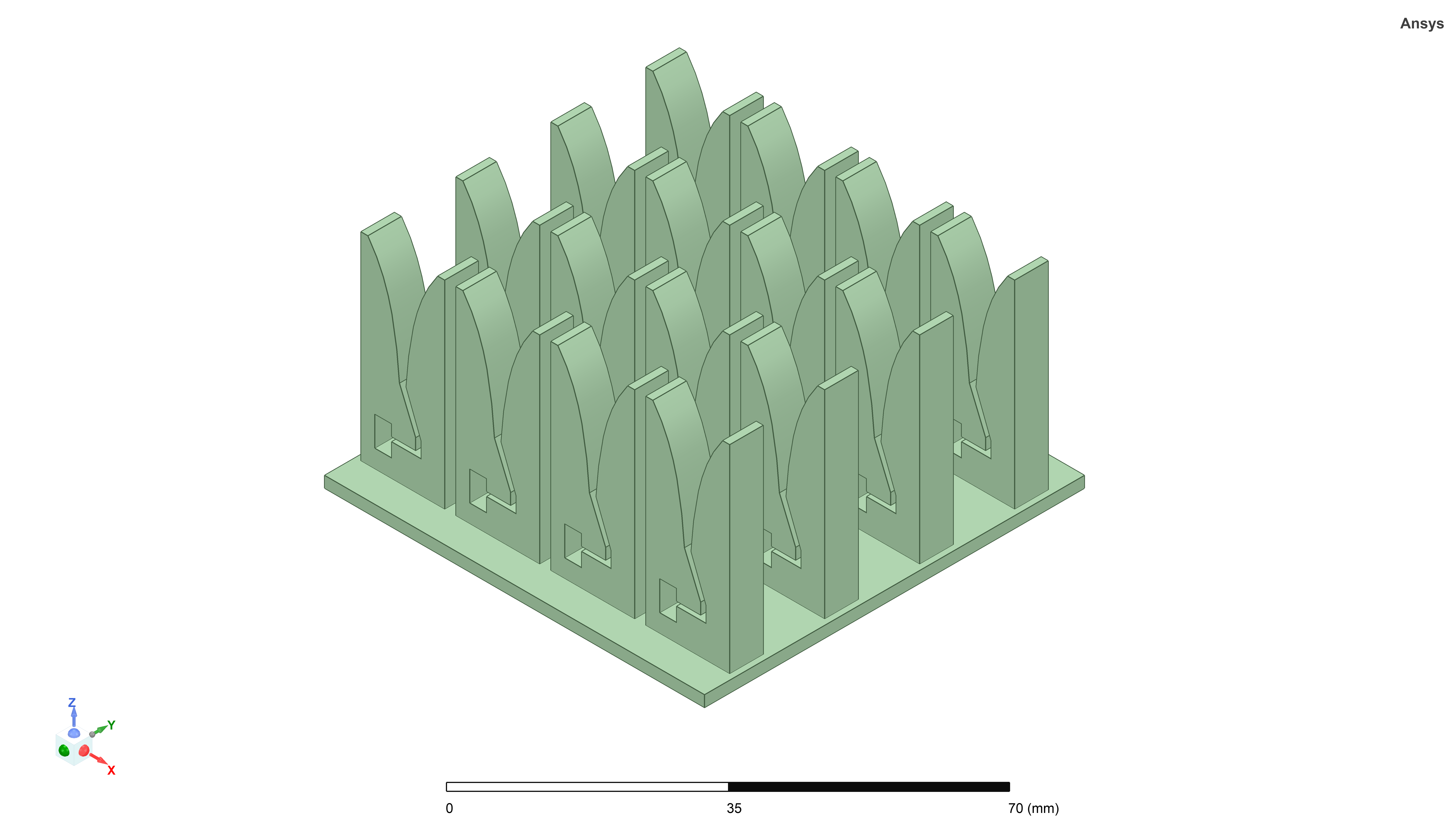}
	
	\captionsetup{
		font=footnotesize, 
		singlelinecheck=false, 
		margin={0pt,0pt} 
	}
	\caption{Structure of a 4×4 SAMV antenna array.}
	\label{fig3}
\end{figure}

\begin{figure}[!t]
	\centering
	$\begin{array}{cc}
		\subfloat[]{\includegraphics[scale=0.3,trim=12cm 2cm 12cm 4.5cm,clip=true]{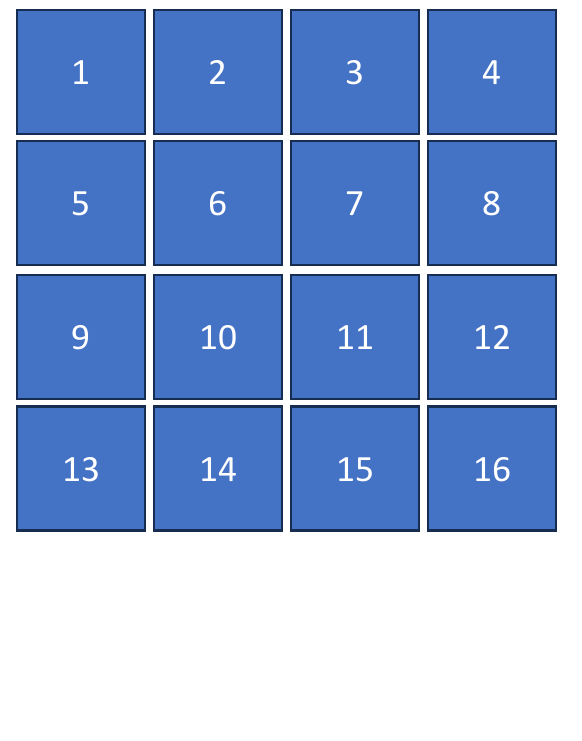}}
		
		\subfloat[]{\includegraphics[scale=0.4,trim=3.5cm 8cm 4.5cm 8cm,clip=true]{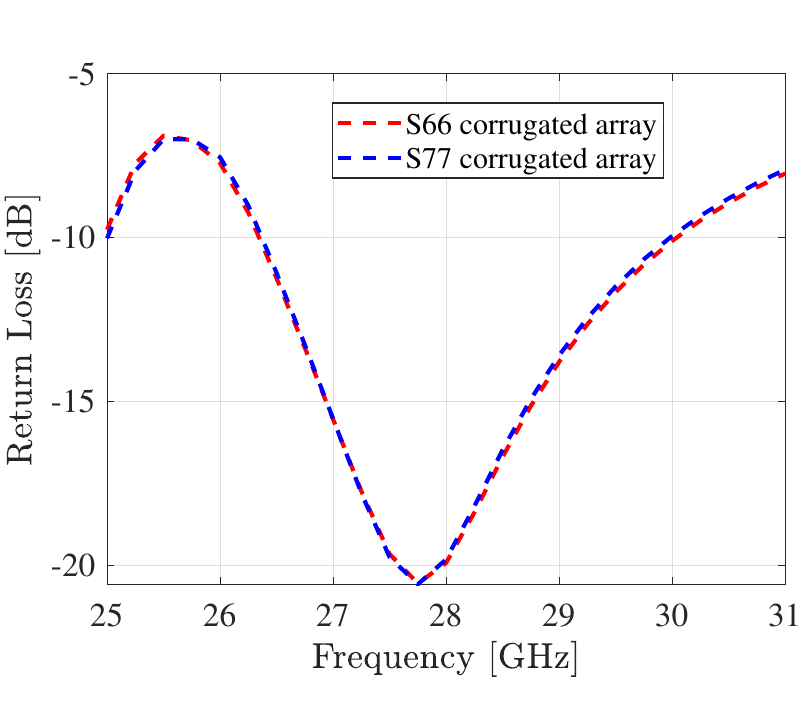}}
		
	\end{array}$
	\captionsetup{
		font=footnotesize, 
		singlelinecheck=false, 
		margin={0pt,0pt} 
	}
	\caption{(a) The arrangement of elements in a 4×4 array configuration. (b) Results of return loss for elements 6 and 7.}
	\label{fig4}
\end{figure}

Next, to achieve the necessary gain for the practical 5G communication applications, we design a Vivaldi antenna array composed of a 16-element ($4 \times 4$) 2-D phased array. Fig.~\ref{fig3} illustrates the 3D view of this Vivaldi antenna array. The arrangement of elements in a 4×4 array in the practical configuration is shown in Fig.~\ref{fig4}(a).

\begin{figure}[!t]
	\centering
	
	\includegraphics[scale=0.4,trim=0cm 8cm 0cm 8cm,clip=true]{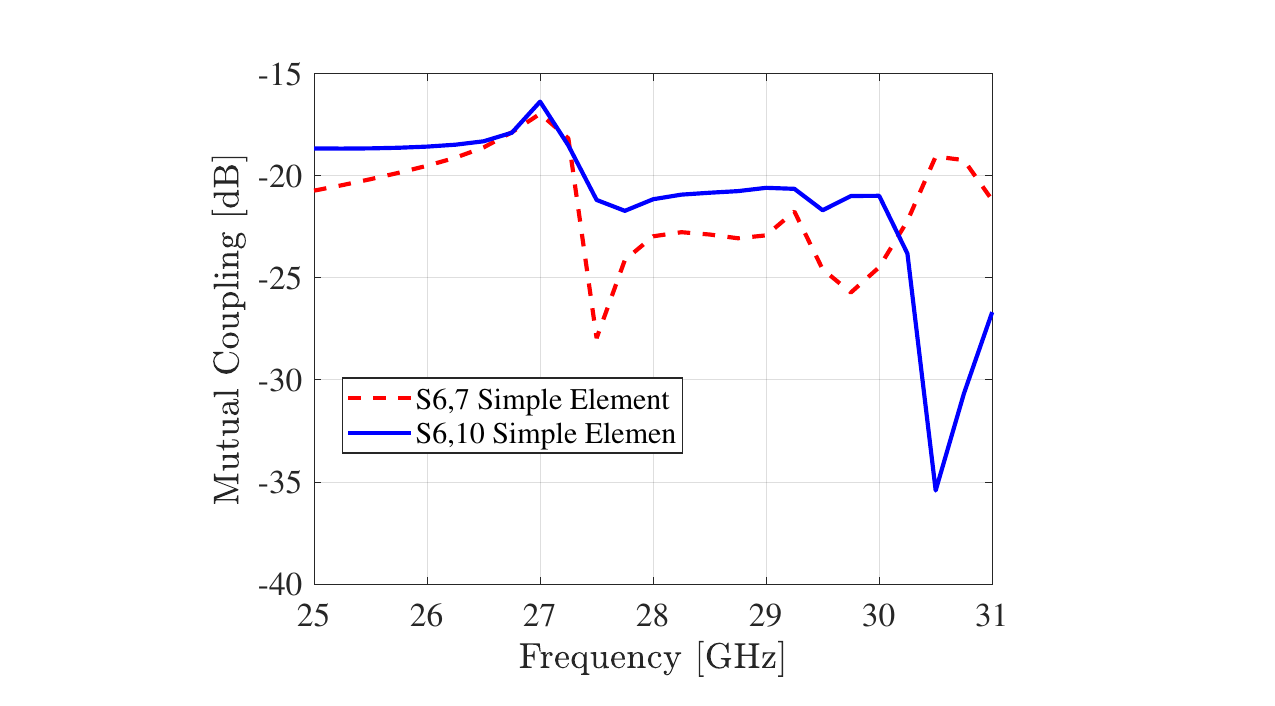}
	\captionsetup{
		font=footnotesize, 
		singlelinecheck=false, 
		margin={0pt,0pt} 
	}
	\caption{Results of mutual coupling parameters for the central elements of the structure.}
	\label{fig5}
\end{figure}

The return loss for elements 6, 7, 10, and 11 is depicted in Fig.~\ref{fig4}(b), which shows the desired 28 GHz resonant frequency of unit elements remained unchanged with good accuracy. In this array configuration, the four central Vivaldi antenna elements have the most effect on the mutual coupling and return loss. Therefore, Fig.~\ref{fig5} shows the mutual coupling parameter for the central element of the 6\textsuperscript{th} position with the 7\textsuperscript{th}  and 10\textsuperscript{th} antenna unit elements in the frequency range of 25-31 GHz. The obtained results for the array antenna are acceptable for many practical applications. However, there is some space for further improvements. 

\begin{figure}[!t]
	\centering
	
	\begin{tikzpicture}
		\node[anchor=south west, inner sep=0] at (0,0) {\includegraphics[scale=0.055,trim=60cm 50cm 60cm 20cm,clip=true]{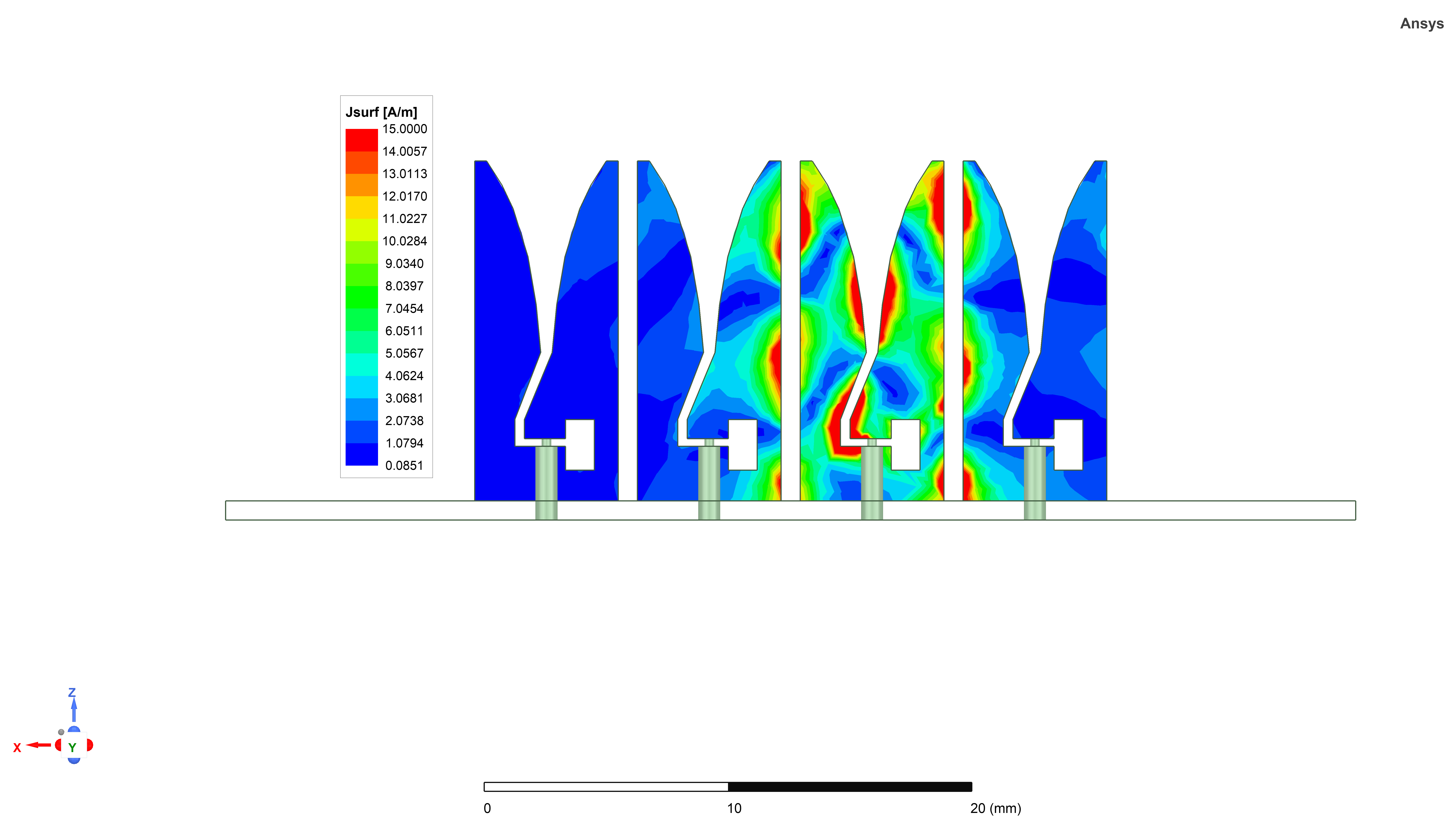}};
		\node at (7.3,4.3) {\textbf{5}}; 
		\node at (5.6,4.3) {\textbf{6}};
		\node at (3.9,4.3) {\textbf{7}};
		\node at (2.3,4.3) {\textbf{8}};
	\end{tikzpicture}
	\hspace{1cm} 

	\captionsetup{
		font=footnotesize, 
		singlelinecheck=false, 
		margin={0pt,0pt} 
	}
	\caption{The SAMV antenna array cross-sectional surface current distribution for the elements of 5-8.}
	\label{fig6}
\end{figure}

To investigate the possible methods for improving the antenna performance, the cross-sectional surface current distribution of the elements 5-8 is illustrated in Fig.~\ref{fig6}. This figure shows how the excitation current of element 6\textsuperscript{th} of the antenna array causes surface current to flow, and its impact on adjacent elements is observable. As a result, these surface currents degrade the return losses and mutual coupling of the four central elements. In the next section, we propose a method for decreasing this surface current and increasing the antenna efficiency.

\begin{figure}[!t]
	\centering
	$\begin{array}{cc}
		
		{\includegraphics[scale=0.45,trim=10.5cm 2cm 14cm 0cm,clip=true]{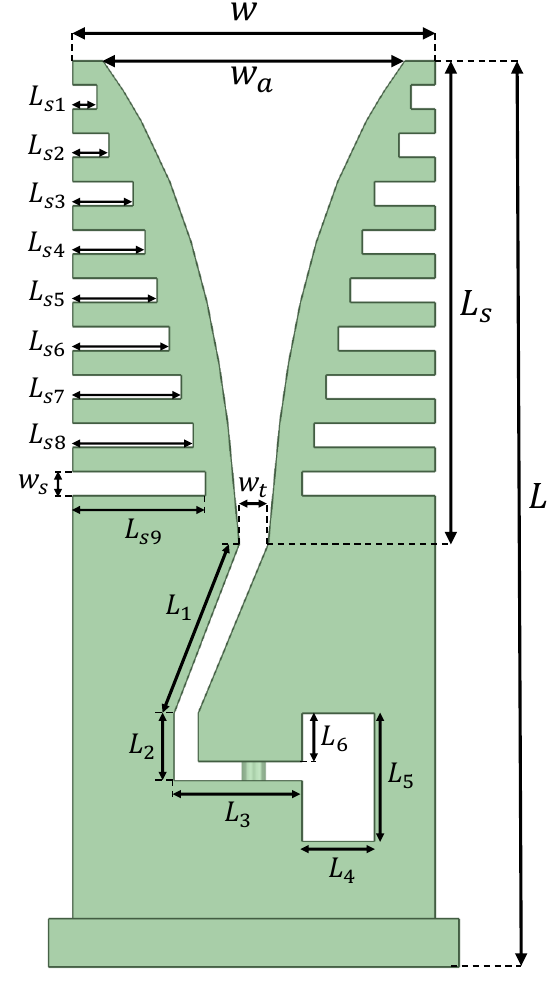}}
		
		{\includegraphics[scale=0.45,trim=10.5cm 1cm 14cm 0cm,clip=true]{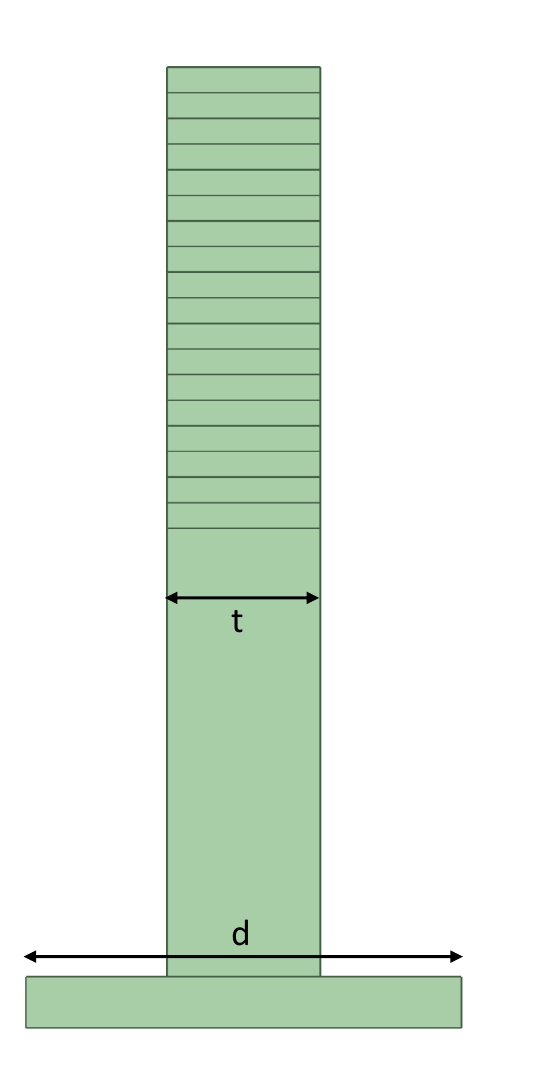}}
		
	\end{array}$
	\captionsetup{
		font=footnotesize, 
		singlelinecheck=false, 
		margin={0pt,0pt} 
	}
	\caption{Corrugated all-metal Vivaldi (CAMV) antenna structure.}
	\label{fig7}
\end{figure}

\section{Corrugated  Unit Element Design}
The presence of surface currents from the antenna aperture towards the ports of the SAMV structure presented in the previous section degrades the return loss within the array structure and reduces the maximum achievable antenna array bandwidth. Therefore, the broad bandwidth capability of the Vivaldi antenna is not fully utilized in the configuration of Fig. 1. To overcome this issue, the antenna must be carefully designed to minimize surface currents. In addition, it is necessary to optimize the antenna geometrical parameters to achieve optimal bandwidth performance.

\begin{table}[tb]
	\captionsetup{
		font=footnotesize 
	}
	\caption{CAMV antenna unit element geometrical parameters.}
	\begin{center}
		\begin{tabular}{@{}ccccl@{}}
			\toprule
			\multicolumn{1}{l}{Parameter} & \multicolumn{1}{l}{Value (mm)} & \multicolumn{1}{l}{Parameter} & \multicolumn{1}{l}{Value (mm)} &  \\ \midrule
			$W$                           & 5.7                           & $L_{s1}$                      & 1                             &  \\
			$W_a$                         & 4.75                          & $L_{s2}$                      & 1.5                           &  \\
			$L_s$                         & 7.6                           & $L_{s3}$                      & 2.5                           &  \\ 
			$L$                           & 14.25                         & $L_{s4}$                      & 3                             &  \\
			$W_t$                         & 0.45                          & $L_{s5}$                      & 3.5                           &  \\
			$W_s$                         & 0.4                           & $L_{s6}$                      & 4                             &  \\
			$L_1$                         & 2.66                          & $L_{s7}$                      & 4.5                           &  \\
			$L_2$                         & 1                             & $L_{s8}$                      & 5                             &  \\
			$L_3$                         & 2                             & $L_{s9}$                      & 5.5                           &  \\
			$L_4$                         & 1.15                          & $d$                           & 6.46                          &  \\
			$L_5$                         & 2                             & $t$                           & 2.28                          &  \\
			$L_6$                         & 0.8                             &  
			$h$                           & 2                             &  \\
			\bottomrule
		\end{tabular}
	\end{center}
	\label{Table1}
\end{table}

\begin{figure}[!t]
	\centering
	
	\includegraphics[scale=0.4,trim=0cm 8cm 0cm 8cm,clip=true]{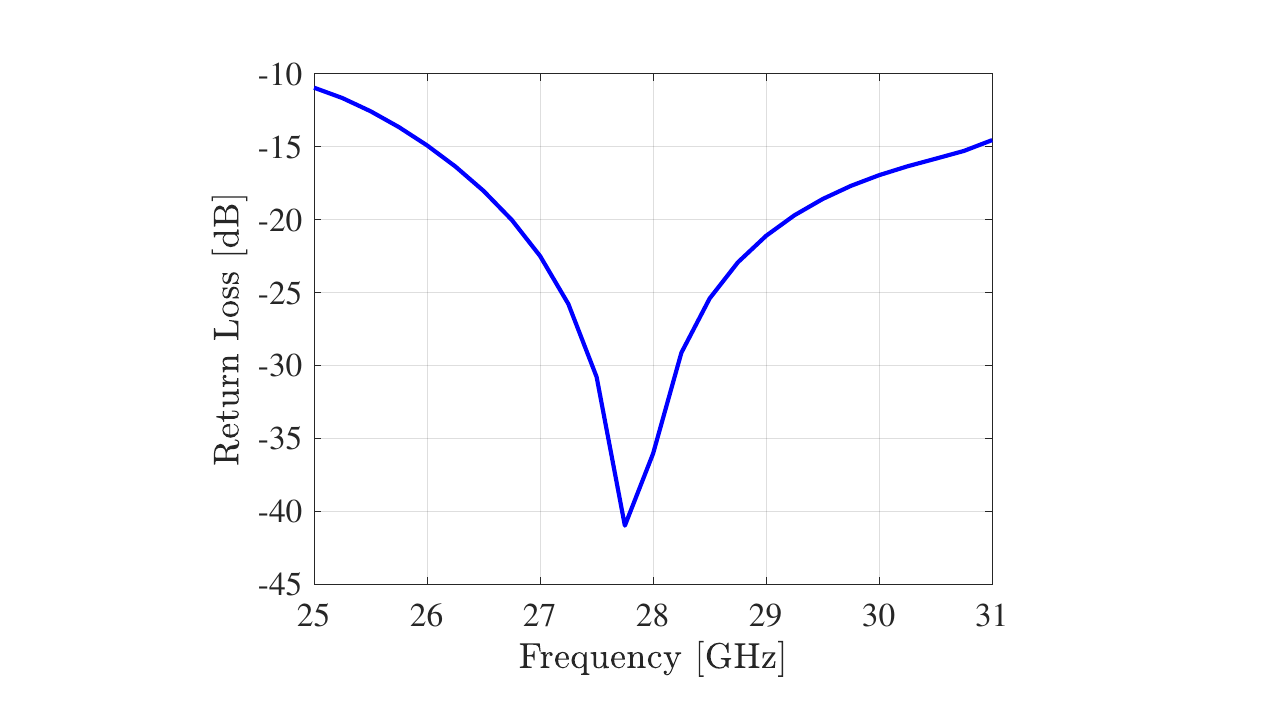}
	\captionsetup{
		font=footnotesize, 
		singlelinecheck=false, 
		margin={0pt,0pt} 
	}
	\caption{Return loss of the corrugated all-metal Vivaldi (CAMV) antenna structure.}
	\label{fig8}
\end{figure}

\begin{figure}[!t]
	\centering
	
	\includegraphics[scale=0.35,trim=4cm 2cm 4cm 1cm,clip=true]{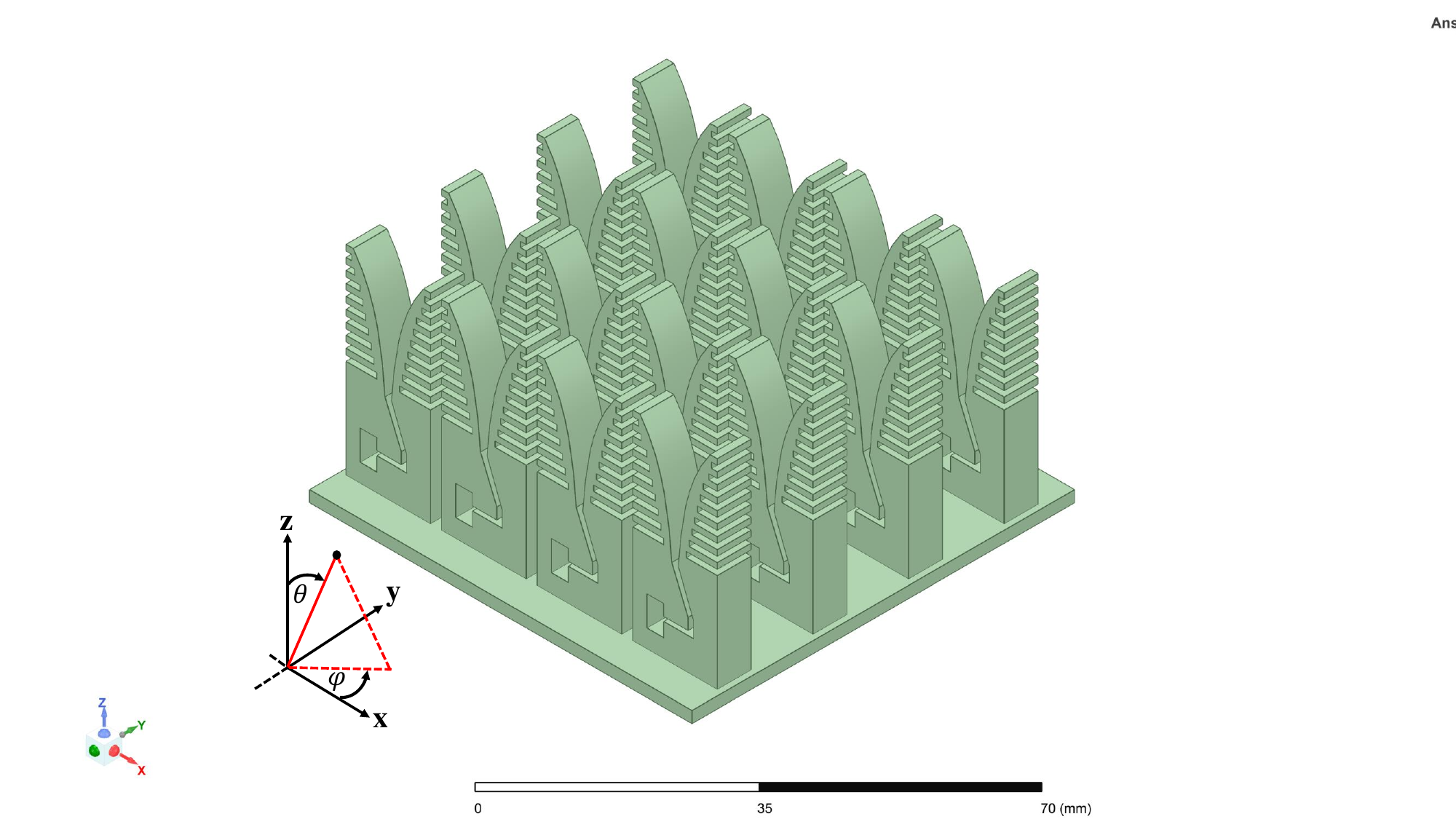}
	\captionsetup{
		font=footnotesize, 
		singlelinecheck=false, 
		margin={0pt,0pt} 
	}
	\caption{Structure of a 4$\times$4 CAMV antenna array.}
	\label{fig9}
\end{figure}

Here, we propose the use of an elaborate corrugation on the edge of the antenna aperture slot. The general structure of the proposed corrugated all-metal Vivaldi (CAMV) antenna unit element is depicted in Fig. 7. The structural adjustments have been implemented to enhance the electrical and radiation characteristics of the antenna unit element at lower operational frequencies. The values of geometrical parameters are listed in Table~\ref{Table1}. 

The simulated return loss of the CAMV antenna unit element is shown in Fig. 8. Considering the -15 dB value as an acceptable return loss in the operational bandwidth, it can be observed a considerable bandwidth improvement in the CAMV antenna in comparison to the SAMV structure of Fig.~\ref{fig2}. Moreover, the main advantage of the CAMV structure show itself when utilized in an array arrangement. 

\begin{figure}[!t]
	\centering
	$\begin{array}{cc}
		
		\subfloat[]{\includegraphics[scale=0.3,trim=3.5cm 8cm 3cm 9cm,clip=true]{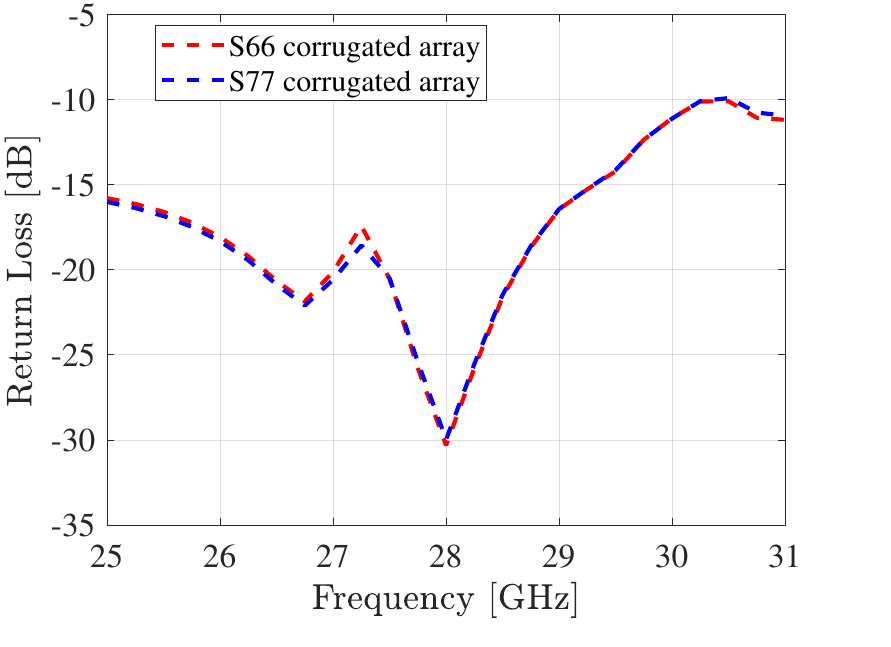}}
		
		\subfloat[]{\includegraphics[scale=0.3,trim=3.5cm 8cm 3cm 9cm,clip=true]{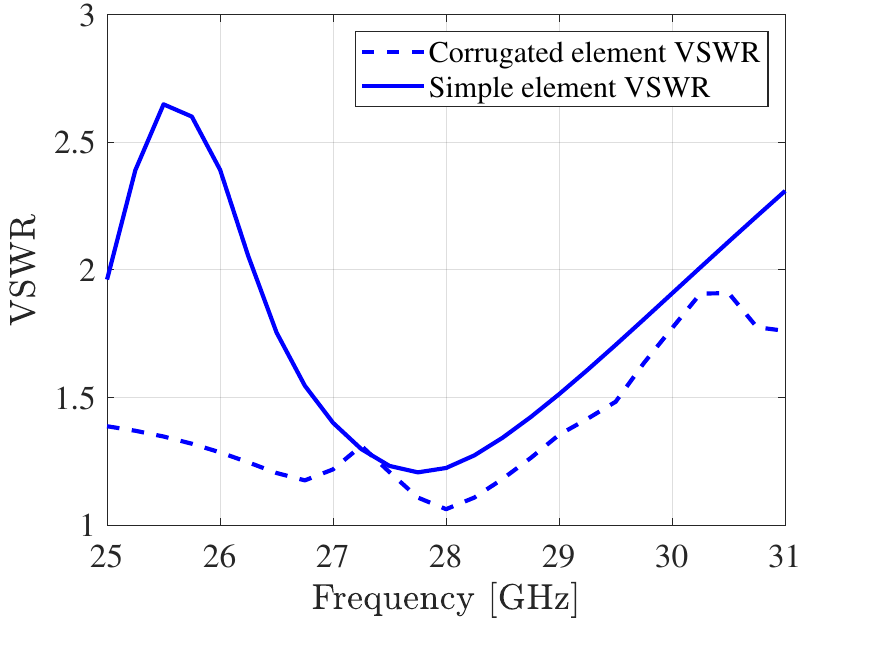}}
		
	\end{array}$
	\captionsetup{
		font=footnotesize, 
		singlelinecheck=false, 
		margin={0pt,0pt} 
	}
	\caption{(a) Results of return loss for elements 6 and 7. (b) Comparison of the 6\textsuperscript{th} antenna element VSWR of the SAMV and CAMV array.}
	\label{fig10}
\end{figure}

\begin{figure}[!t]
	\centering
	
	\includegraphics[scale=0.4,trim=0cm 8cm 0cm 8cm,clip=true]{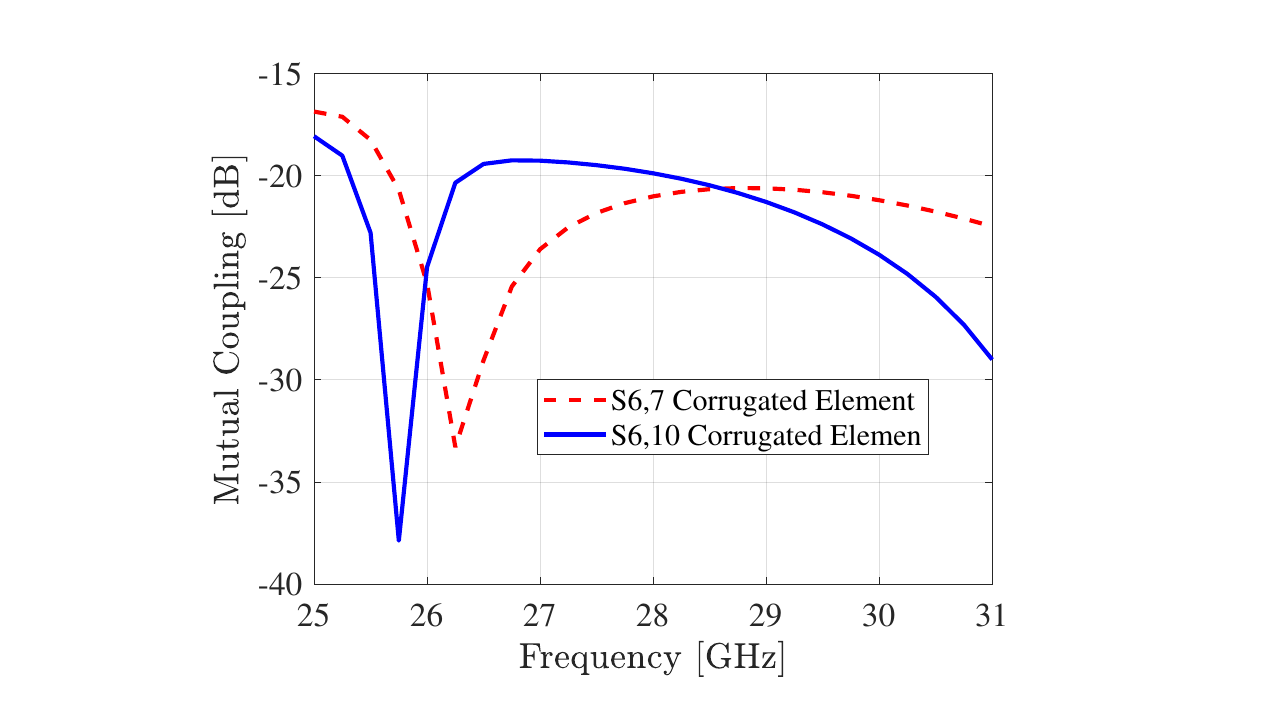}
	\captionsetup{
		font=footnotesize, 
		singlelinecheck=false, 
		margin={0pt,0pt} 
	}
	\caption{Results of mutual coupling parameters for the central elements of the structure.}
	\label{fig11}
\end{figure}

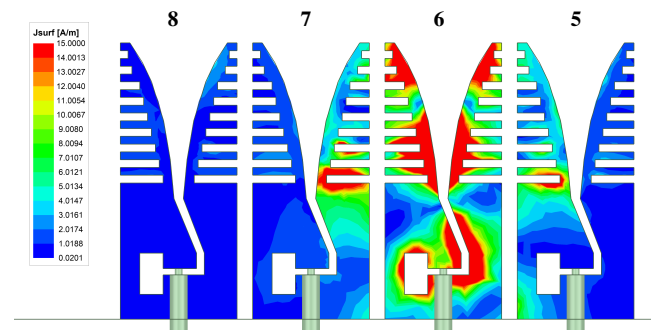
\begin{figure}[!t]
	\centering
	
	\begin{tikzpicture}
		\node[anchor=south west, inner sep=0] at (0,0) {\includegraphics[scale=0.045,trim=35cm 40cm 50cm 20cm,clip=true]{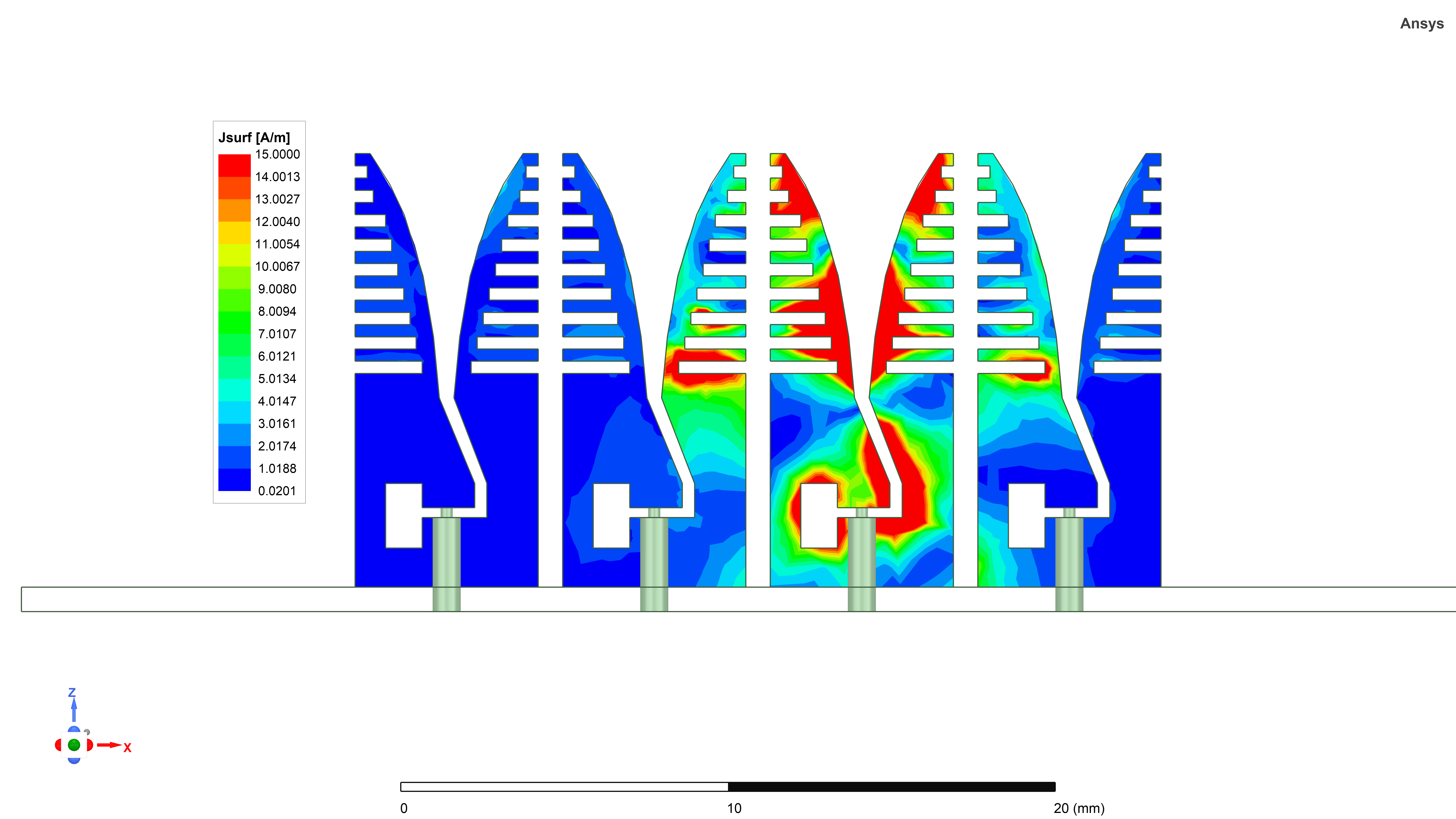}};
		\node at (7.4,4.1) {\textbf{5}}; 
		\node at (5.6,4.1) {\textbf{6}};
		\node at (3.85,4.1) {\textbf{7}};
		\node at (2.1,4.1) {\textbf{8}};
	\end{tikzpicture}
	\hspace{1cm} 

	\captionsetup{
		font=footnotesize, 
		singlelinecheck=false, 
		margin={0pt,0pt} 
	}
	\caption{The CAMV antenna array cross-sectional surface current distribution for the elements of 5-8.}
	\label{fig12}
\end{figure}

The $4 \times 4$ CAMV antenna array configuration is depicted in Fig. 9. First, the simulated return loss for elements 6, 7, 10, and 11 of the CAMV antenna array is shown in Fig.~\ref{fig10}(a). This result shows an improvement in comparison to the return losses of the SAMV antenna array in Fig.~\ref{fig4}(b). More importantly, Fig.~\ref{fig10}(b) compares the VSWR of the selected central antenna unit elements, the 6\textsuperscript{th} element as an example, of the SAMV and CAMV  antenna arrays. There can be observed a significant improvement in the VSWR and, hence, the effective impedance bandwidth of the proposed CAMV phased antenna array. Additionally, Fig.~\ref{fig11} presents the enhanced mutual coupling of the CAMV central element in the 6\textsuperscript{th} position with the 7\textsuperscript{th} and 10\textsuperscript{th} antenna unit elements, in the frequency range of 25-31 GHz, in comparison to the mutual couplings of the SAMV structure.

\begin{figure}[!t]
	\centering
	$\begin{array}{c} 
		
		\subfloat[]{\includegraphics[scale=0.5, trim=3.5cm 8cm 3cm 9cm, clip=true]{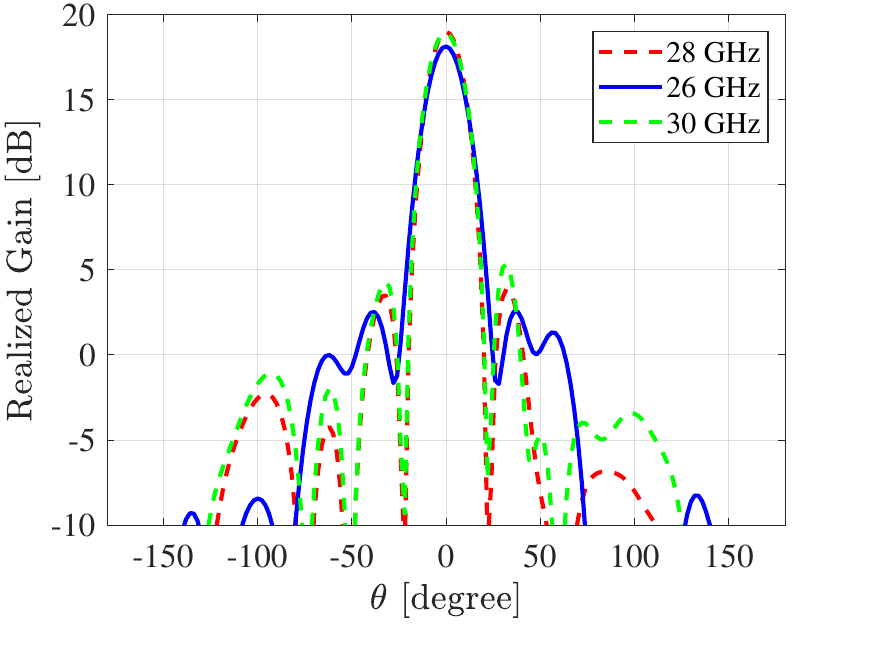}} \\ 
		\subfloat[]{\includegraphics[scale=0.5, trim=3.5cm 8cm 3cm 9cm, clip=true]{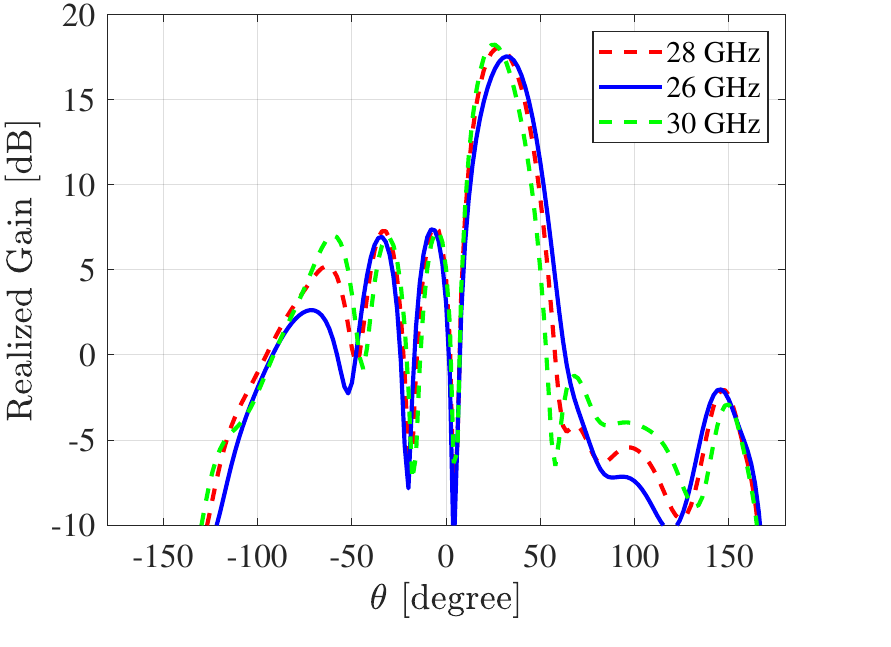}}
		
	\end{array}$
	\captionsetup{
		font=footnotesize, 
		singlelinecheck=false, 
		margin={0pt,0pt} 
	}
	\caption{Realized gain at $\phi =90$ degrees for the frequencies of 26, 28 and 30 GHz; (a) the main beam is at 0 degrees. (b) the main beam is at 30 degrees.}
	\label{fig13}
\end{figure}

The CAMV antenna array cross-sectional surface current distribution for elements 5-8 is shown in Fig.~\ref{fig12}. In the modified corrugated structure, the surface currents are less concentrated at the end of the antenna taper than those of the antenna without corrugation slots. In other words, it can be argued that the slots at the edge of the antenna aperture act as a barrier to prevent current from exiting the antenna and are responsible for the reduction in current flow. Also, the total backscatter leakage of the antenna is reduced as a result of this improved current distribution, which results in an increase in the antenna efficiency. Therefore, The CAMV antenna structure with the corrugated slots has a higher gain, making it more efficient. It also exhibits lower radiation losses, making it more stable and efficient.
 
For beam steering we excited the overall antenna structure and used the following equation to determine the phase distribution between the elements \cite{balanis2016antenna}:

\begin{equation}
	\left\{
	\begin{aligned}
		&\Delta \varphi_x = -\frac{2 \pi d}{\lambda} \sin(\theta) \cos(\varphi)\\
		&\Delta \varphi_y = -\frac{2 \pi d}{\lambda} \sin(\theta) \sin(\varphi) \\
		&\varphi(n, m) = n \cdot \Delta \varphi_x + m \cdot \Delta \varphi_y \\
	\end{aligned}
	\right.
\end{equation}

where \( \Delta_{\phi_x} \), \( \Delta_{\phi_y} \), and \( \phi(n,m) \) are the phase variation along the x-axis, the phase variation along the y-axis, and the phase difference for the element at row \( n \) and column \( m \) in \( xy \)-coordinates in a uniform 2D linear phased array antenna, respectively. In these relations, \( d \) is the distance between elements, \( \lambda \) is the operating wavelength, and also \( \theta \) and \( \phi \) are the elevation and azimuth angles of the main beam direction in standard spherical coordinates. By determining the suitable phase shifts along the x- and y-axis, these equations enable precise beam steering and allow the antenna to focus its signal in a desired direction.

Finally, Fig.~\ref{fig13} illustrates the realized gain of the CAMV antenna array configuration for the frequencies of 26, 28 and 30 GHz when the main beam is at 0 degrees and scanned to 30 degrees. The results of this figure prove the acceptable scanning performance capability of the proposed antenna structure in the desired frequency range.

\balance

\section{CONCLUSION}
In this paper, first, a simple all-metal Vivaldi antenna phased array structure is introduced. The simulated results showed good return loss and mutual coefficients but the maximum expected impedance bandwidth was not achieved. By calculating the surface current distribution, it is proved the surface current leakage of the central elements is the main deconstructive factor for the antenna performance. For this reason, a corrugated all-metal Vivaldi antenna structure is proposed. This configuration led to improvements in the structure’s return losses and bandwidth as a result of decreasing the undesirable leakage surface currents. Additionally, due to its ease of construction, this structure can be a suitable option for fifth-generation communication applications.

\section*{Acknowledgments}
This work was supported by the Royal Society Industry Fellowship, grant number INF-R2-192066, and by the NSF Division of Electrical, Communications, and Cyber Systems (ECCS)- U.K. Engineering and Physical Science Research Council (EPSRC) Award under Grant 2152617.

\bibliographystyle{IEEEtran}
\bibliography{./mybib}

\begin{thebibliography}{10}
\providecommand{\url}[1]{#1}
\csname url@samestyle\endcsname
\providecommand{\newblock}{\relax}
\providecommand{\bibinfo}[2]{#2}
\providecommand{\BIBentrySTDinterwordspacing}{\spaceskip=0pt\relax}
\providecommand{\BIBentryALTinterwordstretchfactor}{4}
\providecommand{\BIBentryALTinterwordspacing}{\spaceskip=\fontdimen2\font plus
\BIBentryALTinterwordstretchfactor\fontdimen3\font minus
  \fontdimen4\font\relax}
\providecommand{\BIBforeignlanguage}[2]{{%
\expandafter\ifx\csname l@#1\endcsname\relax
\typeout{** WARNING: IEEEtran.bst: No hyphenation pattern has been}%
\typeout{** loaded for the language `#1'. Using the pattern for}%
\typeout{** the default language instead.}%
\else
\language=\csname l@#1\endcsname
\fi
#2}}
\providecommand{\BIBdecl}{\relax}
\BIBdecl

\bibitem{yang2018digital}
B.~Yang, Z.~Yu, J.~Lan, R.~Zhang, J.~Zhou, and W.~Hong, ``Digital
  beamforming-based massive mimo transceiver for 5g millimeter-wave
  communications,'' \emph{IEEE Transactions on Microwave Theory and
  Techniques}, vol.~66, no.~7, pp. 3403--3418, 2018.

\bibitem{chettri2019comprehensive}
L.~Chettri and R.~Bera, ``A comprehensive survey on internet of things (iot)
  toward 5g wireless systems,'' \emph{IEEE Internet of Things Journal}, vol.~7,
  no.~1, pp. 16--32, 2019.

\bibitem{ojaroudiparchin20168}
N.~Ojaroudiparchin, M.~Shen, and G.~F. Pedersen, ``8$\times$ 8 planar phased
  array antenna with high efficiency and insensitivity properties for 5g mobile
  base stations,'' in \emph{2016 10th European conference on antennas and
  propagation (EuCAP)}.\hskip 1em plus 0.5em minus 0.4em\relax IEEE, 2016, pp.
  1--5.

\bibitem{wang2022hybrid}
Z.~Wang, Y.~Dong, Z.~Peng, and W.~Hong, ``Hybrid metasurface, dielectric
  resonator, low-cost, wide-angle beam-scanning antenna for 5g base station
  application,'' \emph{IEEE Transactions on Antennas and Propagation}, vol.~70,
  no.~9, pp. 7646--7658, 2022.

\bibitem{kahkonen2020surface}
H.~K{\"a}hk{\"o}nen, J.~Ala-Laurinaho, and V.~Viikari, ``Surface-mounted
  ka-band vivaldi antenna array,'' \emph{IEEE Open Journal of Antennas and
  Propagation}, vol.~2, pp. 126--137, 2020.

\bibitem{rivas2010review}
M.~Rivas, S.~Xie, and D.~Su, ``A review of adaptive beamforming techniques for
  wideband smart antennas,'' in \emph{2010 6th International Conference on
  Wireless Communications Networking and Mobile Computing (WiCOM)}.\hskip 1em
  plus 0.5em minus 0.4em\relax IEEE, 2010, pp. 1--5.

\bibitem{yang2017compact}
B.~Yang, Z.~Yu, Y.~Dong, J.~Zhou, and W.~Hong, ``Compact tapered slot antenna
  array for 5g millimeter-wave massive mimo systems,'' \emph{IEEE Transactions
  on Antennas and Propagation}, vol.~65, no.~12, pp. 6721--6727, 2017.

\bibitem{ibrahim2022slotted}
A.~A. Ibrahim, H.~Zahra, O.~M. Dardeer, N.~Hussain, S.~M. Abbas, and M.~A.
  Abdelghany, ``Slotted antenna array with enhanced radiation characteristics
  for 5g 28 ghz communications,'' \emph{Electronics}, vol.~11, no.~17, p. 2664,
  2022.

\bibitem{sadhu20177}
B.~Sadhu, Y.~Tousi, J.~Hallin, S.~Sahl, S.~Reynolds, {\"O}.~Renstr{\"o}m,
  K.~Sj{\"o}gren, O.~Haapalahti, N.~Mazor, B.~Bokinge \emph{et~al.}, ``7.2 a
  28ghz 32-element phased-array transceiver ic with concurrent dual polarized
  beams and 1.4 degree beam-steering resolution for 5g communication,'' in
  \emph{2017 IEEE International Solid-State Circuits Conference (ISSCC)}.\hskip
  1em plus 0.5em minus 0.4em\relax IEEE, 2017, pp. 128--129.

\bibitem{sun2023reconfigurable}
M.~Sun, H.~Xi, X.~Qi, K.~Xu, H.~Li, Q.~Lv, S.~Hu, S.~Chen, W.~Zhao, X.~Li
  \emph{et~al.}, ``Reconfigurable transmitarray based on frequency selective
  surface for 2d wide-angle beam steering,'' \emph{Electronics}, vol.~12,
  no.~18, p. 3854, 2023.

\bibitem{liu2017ultra}
H.~Liu, Y.~Liu, W.~Zhang, and S.~Gao, ``An ultra-wideband horizontally
  polarized omnidirectional circular connected vivaldi antenna array,''
  \emph{IEEE Transactions on Antennas and Propagation}, vol.~65, no.~8, pp.
  4351--4356, 2017.

\bibitem{gou2022radar}
Y.~Gou, Y.~Chen, and S.~Yang, ``Radar cross section reduction of wideband
  vivaldi antenna arrays with array-level scattering cancellation,'' \emph{IEEE
  Transactions on Antennas and Propagation}, vol.~70, no.~8, pp. 6740--6750,
  2022.

\bibitem{pan2020dual}
Y.~Pan, Y.~Cheng, and Y.~Dong, ``Dual-polarized directive ultrawideband antenna
  integrated with horn and vivaldi array,'' \emph{IEEE Antennas and Wireless
  Propagation Letters}, vol.~20, no.~1, pp. 48--52, 2021.

\bibitem{kahkonen2019dual}
H.~K{\"a}hk{\"o}nen, J.~Ala-Laurinaho, and V.~Viikari, ``Dual-polarized ka-band
  vivaldi antenna array,'' \emph{IEEE Transactions on Antennas and
  Propagation}, vol.~68, no.~4, pp. 2675--2683, 2020.

\bibitem{azari2023super}
A.~Azari, A.~Skrivervik, H.~Aliakbarian, and A.~Sadeghzadeh, ``A super wideband
  dual-polarized vivaldi antenna for 5g mmwave applications,'' \emph{Ieee
  Access}, vol.~11, pp. 80\,761--80\,768, 2023.

\bibitem{kahkonen2022modular}
H.~K{\"a}hk{\"o}nen, J.~Ala-Laurinaho, and V.~Viikari, ``A modular
  dual-polarized ka-band vivaldi antenna array,'' \emph{IEEE Access}, vol.~10,
  pp. 36\,362--36\,372, 2022.

\bibitem{balanis2016antenna}
C.~A. Balanis, \emph{Antenna theory: analysis and design}.\hskip 1em plus 0.5em
  minus 0.4em\relax John wiley \& sons, 2016.

\end{thebibliography}

\profile[MMP]{Mahyar Mehri Pashaki}{was born in Rasht, Iran, in 1995. He received the B.A.Sc. degree (Hons.) in electrical engineering, and the M.Sc. degree from the Department of Electrical Engineering, Sharif University of Technology, in 2022. His current research interests include electromagnetics, microwave engineering, antenna, RF circuits, and periodic structures. He has extensive hands-on experience in both passive and active RF design, including microwave filters, power amplifiers, phased arrays, and millimeter-wave systems.}
\label{profile}

\profile[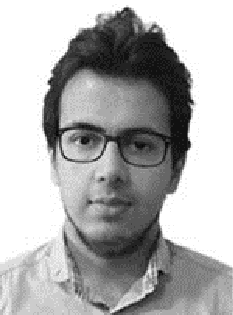]{Mohammad Hossein Koohi Ghamsari}{received his B.Sc. (Hons.) in Electrical Engineering with a focus on Electronics. He earned his M.Sc. in Electrical Engineering, with a focus on Microwave and Photonics, from Sharif University of Technology, Tehran, Iran, in 2022. Following his master's studies, he began collaborating with the Electrical Engineering Department at Sharif University of Technology as a Senior Researcher and Head of a Multidisciplinary Research Group. He is also an active lecturer, delivering university courses and conducting workshops at conferences. He has received several honors and awards, including the Best National Student Book of the Year Award in 2021 (University of Tehran) and an M.Sc. Thesis Fellowship. His research interests include antennas, photonics, metamaterials and metasurfaces, imaging systems, silicon photonics, applied electromagnetics, applied machine learning, and biosensors.}

\profile[AM]{Alireza Mallahzadeh}{(Senior Member, IEEE) received the B.Sc. degree in electrical engineering from the Isfahan University of Technology, Isfahan, Iran, in 1999 and the M.Sc. and Ph.D. degrees in telecommunication engineering from the Iran University of Science and Technology, Tehran, Iran, in 2001 and 2006, respectively.,In 2006, as an Assistant Professor he joined with the telecommunication group of Shahed University, Tehran, Iran. After several years of teaching and extensive research in the field of antennas and passive microwave components, he was promoted to the rank of Associate Professor in 2013.,Dr. Mallahzadeh has acted as a reviewer for reputive journals, such as ieee transaction on antenna and propagation and IEEE Antennas and Wireless Propagation Letters.}

\profile[GG]{Gabriele Gradoni}{received the Ph.D. degree in electromagnetics from University Politecnica delle Marche, Ancona, Italy, in 2010. He was a Visiting Researcher with the Time, Quantum, and Electromagnetics Team, National Physical Laboratory, Teddington, UK, in 2008. From 2010 to 2013, he was a Research Associate with the Institute for Research in Electronics and Applied Physics, University of Maryland, College Park, MD, USA. From 2013 to 2016, he was a Research Fellow with the School of Mathematical Sciences, University of Nottingham, U.K., where he became a Full Professor of Applied Mathematics and Electromagnetics Engineering in 2022. Since May 2023 he has been Full Professor and Chair of Wireless Communications at the 6G Innovation Centre, Institute for Communication Systems, University of Surrey, Guildford, U.K. Since June 2020, he has been an Adjunct Associate Professor with the Department of Electrical and Computer Engineering, University of Illinois at Urbana–Champaign, USA. Since 2020, he has been a Royal Society Industry Fellow at British Telecom, U.K. Since December 2022, he has been a Visiting Fellow with the Department of Computer Science and Technology, University of Cambridge, UK. His research interests include probabilistic and asymptotic methods for propagation in complex wave systems, metasurface modelling, quantum/wave chaos, and quantum computational electromagnetics, with applications to electromagnetic compatibility and modern wireless communication systems. He is a member of the IEEE and the Italian Electromagnetics Society. He received the URSI Commission B Young Scientist Award in 2010 and 2016, the Italian Electromagnetics Society Gaetano Latmiral Prize in 2015, and the Honorable Mention IEEE Transactions on EMC Richard B. Schulz Best Paper Award in 2020. From 2014 to 2021, he was the URSI Commission E Early Career Representative.}

\profile[MKh]{Mohsen Khalily}{(Senior Member, IEEE) is currently a Lecturer in antenna and propagation with the Institute for Communication Systems (ICS), Home of 5G and 6G Innovation Centres, University of Surrey, U.K., where he was a Research Fellow of antennas and propagation, from December 2015 to March 2019. Prior to joining 5GIC, he was a Senior Lecturer with the Wireless Communication Centre (WCC), University Technology Malaysia (UTM).,He has published almost 100 academic articles in international peer-reviewed journals and conference proceedings. His research interests include surface electromagnetic, large intelligent surface, metasurfaces, dielectric resonator antennas, MIMO antennas, phased arrays, circularly polarized antennas for satellite application, hybrid beam-forming, leaky wave antennas, and mm-wave and terahertz antennas and propagation. He is a fellow of the U.K. Higher Education Academy. He is also a member of the IEEE Antennas and Propagation Society, the IEEE Communication Society, and the IEEE Microwave Theory and Techniques Society. He is an Associate Editor of the IEEE Access.}

\end{document}